# Node Disjoint Multipath Routing Considering Link and Node Stability protocol: A characteristic Evaluation

**Dr. Shuchita Upadhayaya and Charu Gandhi**

**Department of Computer Science and Applications, Kurukshetra University, Kurukshetra, Haryana, India**

**Abstract**
Mobile Ad hoc Networks are highly dynamic networks. Quality of Service (QoS) routing in such networks is usually limited by the network breakage due to either node mobility or energy depletion of the mobile nodes. Also, to fulfill certain quality parameters, presence of multiple node-disjoint paths becomes essential. Such paths aid in the optimal traffic distribution and reliability in case of path breakages. Thus, to cater various challenges in QoS routing in Mobile Add hoc Networks, a Node Disjoint Multipath Routing Considering Link and Node Stability (NDMLNR) protocol has been proposed by the authors. The metric used to select the paths takes into account the stability of the nodes and the corresponding links. This paper studies various challenges in the QoS routing and presents the characteristic evaluation of NDMLNR w.r.t various existing protocols in this area.

## 1. Introduction

A Mobile Ad Hoc Networks (MANETs) [1, 2] is collection of mobile/semi mobile nodes with no existing pre-established infrastructure, forming a temporary network. Such networks are characterized by: Dynamic topologies, existence of bandwidth constrained and variable capacity links, energy constrained operations, and highly prone to security threats. Due to all these features routing is a major issue in ad hoc networks. The routing protocols for ad hoc networks have been classified as Proactive/table driven e.g. Destination Sequenced Distance Vector (DSDV) [3], Optimized Link State Routing (OLSR)[4], Reactive/On-demand, e.g. Dynamic Source Routing Protocol (DSR) [5], Ad hoc On-Demand Distance Vector routing protocol (AODV) [6], Temporally Ordered Routing Algorithm (TORA)[4] and Hybrid, e.g. Zone Routing Protocol (ZRP) [7], Hybrid Ad hoc Routing Protocol (HARP) [23].

Quality of Service (QoS) based routing is defined in RFC 2386 [8] as a "Routing mechanism under which paths for flows are determined based on some knowledge of resource availability in the network as well as the QoS requirement of flows." The main objectives of QoS based routing are[8]:Dynamic determination of feasible paths for accommodating the QoS of the given flow under policy constraints such as path cost, provider selection , optimal utilization of resources for improving total network throughput and graceful performance degradation during overload conditions giving better throughput. QoS routing strategies are classified as source routing, distributed routing and hierarchical routing [9].

QoS based routing becomes challenging in MANETs, as nodes should keep an up-to-date information about link status. Also, due to the dynamic nature of MANETs, maintaining the precise link state information is very difficult. Finally, the reserved resource may not be guaranteed because of the mobility-caused path breakage or power depletion of the mobile hosts. QoS routing should rapidly find a feasible new route to recover the service. Our motive in this paper is to design a routing technique, which considers all three above problems together. We define a metric that attempts to maintain a balance between mobility and energy constraints in MANETs. We use Dynamic Source Routing (DSR) [5], as the base protocol to design our model.

## 2. Related Works

In the recent period lot of research has been done in QoS based, multi-path and node disjoint routing. Lately, the upcoming concern is the energy issues in mobile ad hoc networks (MANETs) The recent studies extensively focused on the multipath discovering extension of the on- demand routing protocols in order to alleviate single-path problems like AODV[6 ] and DSR[5], such as high route discovery latency, frequent route discovery attempts and possible improvement of data transfer throughput. The AODVM (AODV Multipath) [10], is a multipath extension to AODV. These provide link-disjoint and loop free paths in AODV. Cross-layered multipath AODV (CM-AODV) [11], selects multiple routes on demand based on the signal-to-interference plus noise ratio (SINR) measured at the physical layer. The Multipath Source Routing (MSR) protocol [12], a multipath extension to DSR, uses weighted round robin packet distribution to improve the delay and throughput. Split Multipath Routing (SMR) [13] is another DSR extension, which selects hop count limited and maximally disjoint multiple routes. Node-Disjoint Multipath Routing (NDMR) [14], provides with node-disjoint multiple paths. Other energy aware multipath protocols which give disjoint paths are Grid-based Energy Aware Node-Disjoint Multipath Routing Algorithm GEANDMRA) [15], Energy Aware Source Routing (EASR) [I6] and Energy Aware Node Disjoint multipath Routing (ENDMR) [I7]. The Lifetime-Aware Multipath Optimized Routing (LAMOR) [18] is based on the lifetime of a node which is related to its residual energy and current traffic conditions. Cost- effective Lifetime Prediction based Routing (CLPR) [19], combines cost efficient and lifetime predictions based routing. Minimum Transmission Power Routing (MTPR) [20], Power-aware Source Routing (PSR) [21].

## 3. QoS Routing Challenges in MANETs

Because of the inherent properties of MANETs, establishing a stable path which can adhere to the QoS requirements is a challenging issue. The stability issues of a data transmission system in a MANET can be studied under following aspects:

1. **Existence of mobile nodes (Mobility factor):** A MANET consists of mobile nodes. Nodes form the network only when they are in the communication range of each other. If they move out of range, link between two nodes is broken. At times, breakage of a single link can lead to the major network partitioning. Hence, mobility of the nodes is a major challenging issue for a stable network. Also, breakdown of certain links results in routing decisions to be made again.
2. **Limited battery /energy factor:** Mobile nodes are battery driven. Thus, the energy resources for such networks are limited. Also, the battery power of a mobile node depletes not only due to data transmission but also because of interference from the neighboring nodes. Thus, a node looses its energy at a specific rate even if it is not transferring any data packet. Hence the lifetime of a network largely depends on the energy levels of its nodes. Higher the energy level, higher is the link stability and hence, network lifetime. Also lower is the routing cost.
3. **Multiple paths:** To send data from a source to destination, a path has to be found before hand. If a single path is established, sending all the traffic on it will deplete all the nodes faster. Also, in case of path failure, alternate path acts as a backup path. Thus, establishing multiple paths aids not only in traffic engineering but also prevents faster network degradation
4. **Node-disjoint paths:** Multiple paths between two nodes can be either link-disjoint or node disjoint. Multiple link-disjoint paths may have one node common among more than one path. Thus, traffic load on this node will be much higher than the other nodes of the paths. As a result, this node tends to die much earlier than the other nodes, leading to the paths to break down much earlier. Thus, the presence of node disjoint paths prolongs the network lifetime by reducing the energy depletion rate of a specific node.

## 4. Problem Issue

From the above mentioned challenges to achieve Quality of Service (QoS) routing in MANETs, it can be concluded that the major reasons for link and hence, path breakage are two fold:

a) Node dying of energy exhaustion
b) Node moving out of the radio range of its neighboring node

Hence, to achieve the route stability in MANETs, both link stability and node stability is essential.
The above mentioned techniques consider either of the two issues. Techniques in [19, 10, 13, and 20] calculate only multiple paths. Both stability issues are neglected in these. The work in [11] measures route quality in terms of SINR, which gives reliable links, but overall networks stability is not considered. Though [19] uses lifetime of a node as a generalized metric, it does not consider the mobility and energy issues which are critical to network - lifetime estimation. The protocol in [17] considers the energy issues in terms of the energy expenditure in data transmission, but the lifetime of the node and mobility factor is not discussed. [7, 15, 16, 21] consider only energy metric to route the traffic.

Also, to send a packet from a source to destination many routes are possible. These routes can be either link disjoint or node-disjoint. Node disjoint protocols have an advantage that they prevent the fast energy drainage of a node which is the member of multiple link disjoint paths [14]. Hence, a technique which finds multiple node-disjoint paths considering both link and node stability has been proposed. The attempt is to find multiple node disjoint routes which consider both link stability and the node stability on their way.

## 5. Metrics Used

To measure link and node stability together we are using two metrics, Link Expiration Time (LET) [19] and Energy Drain Rate (EDR) [22] respectively. These two metrics can be used to generate a composite metric which keeps track of the stability level of the entire path.

Mobility Factor: The mobility factor, Link Expiration Time (LET), was proposed in [19], by using the motion parameters (velocity, direction) of the nodes. It says that if r is the transmission distance between the two nodes, i and j, $(x_i, y_i)$ and $(x_j, y_j)$ be the position co-ordinates and $(v_i, \theta_i)$ and $(v_j, \theta_j)$ be the (velocity, direction) of motion of nodes. LET is defined as:

$$LET = -(ab+cd) + Q/(aPP^2+c^2) \quad (1)$$

Where, $Q = \sqrt{\{(a^2+c^2) r^2 - (ad-bc)^2\}}$ and,
$a = v_i \cos\theta_i - v_j \cos\theta_j$, $b = x_i-x_j$, $c = v_i \sin\theta_i - v_j \sin\theta_j$, and $d = y_i - y_j$

The motion parameters are exchanged among nodes at regular time intervals through GPS. The above parameter suggests that if the two nodes have zero relative velocity, i.e., $v_i = v_j$ and $\theta_j = \theta_j$, the link will remain forever, as LET will be ∞.

Energy factor: Most of the energy based routing algorithms [10, 17, and 21], send large volume of data on the route with maximum energy levels, As a result, nodes with much higher current energy levels will be depleted of their battery power very early. The mobile node also loses some of it energy due to overhearing of the neighboring nodes. Thus, a node is losing its power over a period of time even if no data is being sent through it.

Viewing all these factors a metric called Drain Rate (DR) was proposed in [22], Drain Rate of a node is defined as the rate of dissipation of energy of a node. Every node calculates its total energy consumption every T secs and estimates the DR, Actual Drain Rate is calculated by exponentially averaging the values of $DR_{old}$ and $DR_{new}$ as follows:

$$DR_i = \alpha DR_{old} + (1-\alpha) DR_{new} \quad (2)$$

Where, $0 < \alpha < 1$, can be selected so as to give higher priority to updated information. Thus, higher the Drain Rate, faster the node is depleted of its energy.

## 6. Node Disjoint Multipath Routing Considering Link and Node Stability (NDMLNR)

The main aim of the proposed work is to find the multiple node disjoint routes from source to a given destination Also it keeps track of the route bandwidth which can be further used by the source to select the optimal routes. From the factors Link Expiration Time (LET) [19] and Drain Rate (DR) [22] it is inferred that the Link Stability:

a) Depends directly on Mobility factor

b) Depends inversely on the energy factor

Hence, Link Stability Degree (LSD) is defined as:

$$LSD = \text{Mobility factor} / \text{Energy factor} \quad (3)$$

It defines the degree of the stability of the link. Higher the value of LSD, higher is the stability of the link and greater is the duration of its existence. Thus, a route having all the links with $LSD > LSD_{thr}$ is the feasible.

We choose the Dynamic Source Routing (DSR) [5] protocol as a candidate protocol. Modifications are made to the Route Request (RREQ) and Route Reply (RREP) packets to enable the discovery of link stable node disjoint paths. The proposed scheme has three phases: Route Discovery, Route Selection and Route Maintenance. The various phases are described as follows:

6.1 Route Discovery

The source node when needs to send packet to some destination node, starts the route discovery procedure by sending the Route Request packet to all its neighbors .In this strategy, the source is not allowed to maintain route cache for a long time, as network conditions change very frequently in terms of position and energy levels of the nodes. Thus, when a node needs route to the destination, it initiates a Route Request packet, which is broadcasted to all the neighbors which satisfy the broadcasting condition. Route Request packet of NDMLNR is shown in figure 1.

| SA | DA | Type | ID | TTL | Hops | Bandwidth | LSD | Path | Velocity | Direction | Position |
|----|----|------|----|----|------|-----------|-----|------|----------|-----------|----------|

Figure 1. RREQ packet

**Type (T) field**: It indicates the type of packet.
**SA (Source Address) field**: It carries the source address of node.
**ID field**: unique identification number generated by source to identify the packet.
**DA (Destination Address) field**: It carries the destination address of node.
**Time to Live (TTL) field**: It is used to limit the life time of packet, initially, by default it contains zero.
**Hop field**: It carries the hop count; the value of hop count is incremented by one for each node through which packet passes. Initially, by default this field contains zero value.
**LSD field**: when packet passes through a node, its LSD value with the node from which it has received this packet is updated in the LSD field. Initially, by default this field contains zero value.
**Bandwidth field**: carries the cumulative bandwidth of the links through which it passes; initially, by default this field contains zero value.
**Path field**: It carries the path accumulations, when packet passes through a node; its address is appended at end of this field.

The node's current velocity, direction and position are updated at each node in the respective fields before forwarding the RREQ packet..

Every node maintains a Neighbor Information Table (NIT), to keep track of multiple RREQs. With following entries Source Address, Destination Address, Hops, LSD, ID and bandwidth.

| SA | DA | ID | Hops | LSD | Bandwidth |
|----|----|----|----|-----|-----------|

Figure 2. Neighbor Information Table (NIT)

As RREQ reaches a node it enters its information in the NIT. It makes all the entries for the requests till Wait Period. At the end of the Wait Period, it accepts the request with the highest value in LSD field. It adds the value of the link bandwidth to the Bandwidth field of the RREQ packet. If two RREQs have same LSD values, the one with lesser value of hop count is selected. In case, hops are also same, one with higher bandwidth is selected. In the worst case, RREQ is selected on First-come-first -serve basis. This prevents loops and unnecessary flooding of RREQ packets. None of the intermediate nodes is allowed to send RREP if it has the current route to the destination. As doing this may lead to those paths which do not fulfill current QoS requirements.

The route discovery and selection process is described in figure 3. Details of this technique have been described by the authors in [24].

6.2 Route Maintenance

In case, LSD of a node falls below $LSD_{thr}$, it informs its predecessor node of the node failure by sending the NODEOFF message. Once a node receives such a message, it sends the ROUTEDISABLE message to the source node. Source can then reroute the packets to the backup routes. If no backup route exists, the source then starts the route discovery procedure again. An illustration of this technique with an example has been presented by the authors in [24].

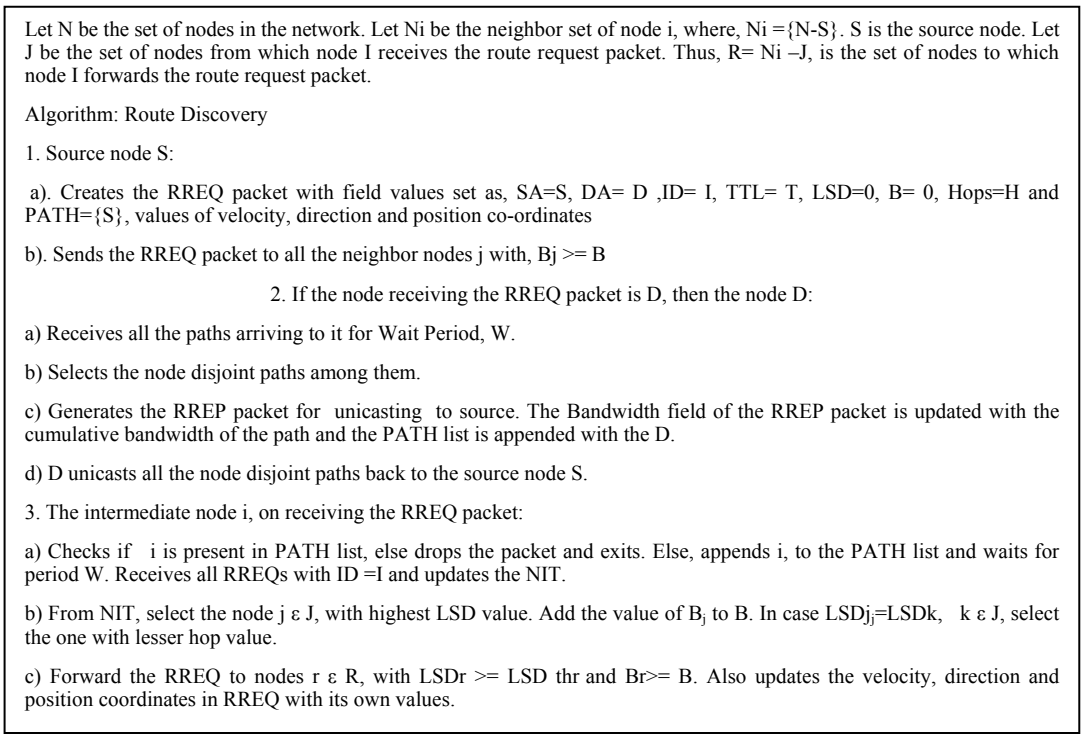

Let N be the set of nodes in the network. Let Ni be the neighbor set of node i, where, Ni ={N-S}. S is the source node. Let J be the set of nodes from which node I receives the route request packet. Thus, R= Ni –J, is the set of nodes to which node I forwards the route request packet.

Algorithm: Route Discovery

1. Source node S:

 a). Creates the RREQ packet with field values set as, SA=S, DA= D ,ID= I, TTL= T, LSD=0, B= 0, Hops=H and PATH={S}, values of velocity, direction and position co-ordinates

b). Sends the RREQ packet to all the neighbor nodes j with, $B_j >= B$

2. If the node receiving the RREQ packet is D, then the node D:

a) Receives all the paths arriving to it for Wait Period, W.

b) Selects the node disjoint paths among them.

c) Generates the RREP packet for unicasting to source. The Bandwidth field of the RREP packet is updated with the cumulative bandwidth of the path and the PATH list is appended with the D.

d) D unicasts all the node disjoint paths back to the source node S.

3. The intermediate node i, on receiving the RREQ packet:

a) Checks if i is present in PATH list, else drops the packet and exits. Else, appends i, to the PATH list and waits for period W. Receives all RREQs with ID =I and updates the NIT.

b) From NIT, select the node $j \epsilon J$, with highest LSD value. Add the value of $B_j$ to B. In case $LSD_{j} = LSD_k$, $k \epsilon J$, select the one with lesser hop value.

c) Forward the RREQ to nodes $r \epsilon R$, with $LSD_r >= LSD\ thr$ and $B_r >= B$. Also updates the velocity, direction and position coordinates in RREQ with its own values.

Figure3. Route discovery and selection process

## 7. Example

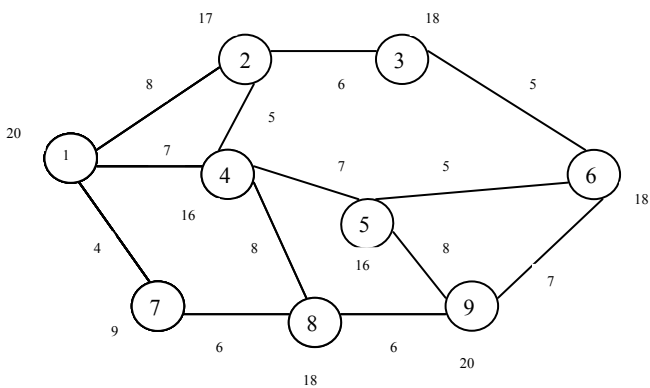

Figure 4. An example network

Let us illustrate our technique with the following example network shown in figure 3. Suppose node 1 is the source node and node 6 is the destination. Let $LSD_{thr}$ equals to 15. Let B equals to 5 mbps. To send the packet, node 1 checks its neighbors (2.4.7) for their LSD value Out of these node 7 has value 9<15. So, node 1 sends the packets only to nodes 2 and 4. Node 2 receives this packet for the first time, makes entry in its NIT for the RREQ packet as (1, 6, 1, 1, 20, and 8) and starts Wait Time, 5 secs here. Node 2 now checks its neighbors, updates the path field as,1-2 and the bandwidth field to 8 and forwards RREQ to both 4 and 3. At node 4, it may receive two RREQ packets during Wait Time. One from node 1 directly, and, the other via node 2. It has two entries in its NIT (1,6,1,1.20,8) and (1,6,1,2,17,13). At this moment it selects the one from node 1 with higher LSD value, 20. It updates the path field of the RREQ packet as 1-4 and the bandwidth field to 7. It forwards the packet to both its neighbors, 5 and 8, with LSD values 16 and 18 respectively. Node 3 has only one neighbor, 6 which satisfies the LSD value and hence, it updates RREQ path field as 1-2-3 and the bandwidth field to 14 and forwards the packet to node 6. Node 6 now receives a path from source node 1. It appends its own ID to it. Thus, first path is 1-2-3-6 and bandwidth of this path is 17. Node 5 after receiving the RREQ packet with path 1-4, checks for its neighbors and forwards RREQ with updated path field to 1-4-5 and bandwidth field to14 to nodes 9 and 6 Node 6 now receives another path,1-4-5.It appends its ID to it, to get the path, 1-4-5-6 with bandwidth 19. Node 8 after receiving the RREQ packet forwards it to its neighbor, 9, after updating path field to 1-4-8 and bandwidth field to 15 Node 9 can receive two packets in its wait time, one from node 5 and the other from node 8. It updates its NIT as (1,6,1,3,16,22) and (1,6,1,3,18,21). To select from the one, it chooses one from node 8 as its LSD value is higher, 18. It then forwards the request after updating the path field as 1-4-8-9 and bandwidth field to 21. Node 6 again receives another path 1-4-8-9.It appends its ID to this path to get 1-4-8-9-6 with bandwidth 28.Now node 6 receives two paths 1-4-5-6 and 1-4-8-9-6 with node 4 as common node. It selects the one with higher bandwidth i.e. Path, 1-4-8-9-6 with bandwidth 28.

In the next section we present the characteristic evaluation of our technique w.r.t various protocols which attempt to achieve QoS routing considering the stability of the network in one way or the other.

## 8. Characteristic Evaluations

To evaluate our protocol we consider the following protocols: Traffic load and lifetime Deviation based Power-aware Routing protocol (TDPR) [26], Cost-effective Lifetime Prediction based Routing (CLPR) [19], Energy aware Node Disjoint Multipath Routing (ENDMR) [17], QoS Aware Stable and Effective Lifetime Prediction Routing (QSEL)[27], Collision Constrained Energy Algorithm (ECCA)[25]. Each of these protocols, attempt to provide QoS routing in their own manner. We evaluate these on the factors: Mobility issue, energy factors, multiple paths and node-disjoint routes.

The ECCA [25] provides for multiple node disjoint paths between source and destination. It calculates the total transmission power needed for transmission. It also deals with the problem of node interference and uses a correlation factor to select the paths that have minimum probability of collisions and thus prevents a node from dying of over burden. Hence, it attempts to provide energy saving at the nodes. But it does not take into account the power status of the nodes and thus, for how long the network will remain stable. Also the position of the node is used only to calculate the transmission power. Link stability due to its movement is not considered which will affect the transmission power calculation, which is a very lengthy procedure. Thus, recalculating new paths in case of path failure becomes tedious and introduces delay.

The QSEL [27] selects the stable paths. It calculates the lifetime of the path by using the location predictions, Link Expiration Time and the communication cost. The communication cost is also only predicted. No exact parameter is used to calculate it. Also it does not consider the issue of multiple paths if a node moves out of transmission range. The energy factor is also absent while selecting the paths. The CLPR [19] on the similar grounds calculates the predicted lifetime from the residual energy and rate of depletion of energy per packet at a particular node. But it does not consider the mobility factor, which is the critical issue in MANETs. Though both these protocols attempt to find longer lifetime and least cost paths, they select a single path and the entire route discovery is to be made in case of a node/path failure.

The TDPR [26] protocol uses node lifetime prediction function. It considers not only residual battery capacity and transmission power but also the traffic load. The traffic load is defined as the total amount of the expected energy consumption by the active paths of the node. The transmission power of each route is stored in the routing table. This does not take into account the route breakages due to node movement. Also, multiple paths are not calculated.

The ENDMR [17] is a node disjoint multipath routing protocol. It assigns the cost to the node based on its residual energy. The routing process is such that it limits the route request packet broadcast and hence, prevents loop formation. The paths are selected which have minimum cost and maximum routing energy. Though this protocol considers all the factors, still the node mobility and network stability is not considered. Table 1 summarizes the characteristic comparison of these protocols with NDMLNR.

## 9. Conclusions

From table 1 it can be inferred that majority of the techniques consider one factor or another to establish QoS paths. But to fulfill all the challenges posed by routing conditions in a MANET our protocol ranks much higher than the cases studied so far, as it attempts to cater all the challenges encountered so far in QoS routing in MANETs. It is expected that the experimental evaluation of this technique will further prove it to be more efficient than the fellow techniques.

| Protocol | Mobility Factor | Energy factor | Multiple Paths | Node-disjoint paths |
|---|---|---|---|---|
| NDMLNR | Y | Y | Y | Y |
| ECCA | N | Only Transmission power. Not the individual node energy | Y | Y |
| QSEL | Y | N | N | N |
| CLPR | N | Y Predicts lifetime on basis of energy consumed per packet | N | N |
| TDPR | N | Y Residual battery power and transmission power is used to calculate the traffic load | N | N |
| ENDMR | N | Y Residual battery capacity is used as cost function | Y | Y |

Table 1. Characteristic Evaluation of NDMLNR

**Dr. Shuchita Upadhayaya** : She did her Masters of Computer Applications (M.C.A) in 1987 and Ph.D in Computer Science in 2003. She worked as Lecturer at Banasthali Vidyapith, Rajasthan, India, for nearly 2 years. After that she is serving as Reader in the Department of Computer Science and Applications, Kurukshetra University, Kurkshetra, India. She has teaching and research experience in post graduate courses of above 21 years. She has above 35 publications in various national and international conferences and international journals to her credit. Her research area is Routing in Computer Networks. Her areas of interests include: Data Communications and Computer Networks and Computer Graphics. She is reviewer of various known national journals. She is the member of Indian Society of Information Theory and Applications and International Forum for Interdisciplinary Mathematics.

**Charu Gandhi**: She did her Bachelor of Technology in Computer Science in 2003, Master of Technology in Computer Science in 2005. She has worked as Lecturer at National Institute of Technology, Kurukshetra , for 1 year, Senior Lecturer at Institute of Technology and Management, Gurgaon, India, for 4 years. Presently, she is a Research Scholar at Department of Computer Science and Applications, Kurukshetra University, Kurukshetra, India. She has an experience of 5 years in teaching and research. Her research area is Quality of Service Routing in Mobile Ad hoc Networks. Her areas of interests include: Computer Networks, Mobile communications and Data Structures. She is a member of Indian Society of Technical Education and ACM.